# Electric Vehicle Attack Impact on Power Grid Operation

Mohammad Ali Sayed [a], Ribal Atallah [b], Chadi Assi [a,*], Mourad Debbabi [a]
a: Concordia Institute for Information Systems Engineering, Concordia University, Montreal, Quebec, Canada
b: Hydro-Quebec Research Institute, Montreal, Quebec, Canada
*Corresponding Author. Email: chadi.assi@concordia.ca

**Abstract**

The increasing need for reducing greenhouse gas emissions and the drive for green cities have promoted the use of electric vehicles due to their environmental benefits. In fact, countries have set their own targets and are offering incentives for people to purchase EVs as opposed to traditional gasoline-powered cars. Manufacturers have been hastily deploying charging stations to meet the charging requirements of the EVs on the road. This rapid deployment has contributed to the EV ecosystem's lack of proper security measures, raising multiple questions related to the power grid security and vulnerability. In this paper, we offer a complete examination of the EV ecosystem from the vulnerability to the attacks and finally the solutions. We start by examining the existing vulnerabilities in the EV ecosystem that can be exploited to control the EV charging and launch attacks against the power grid. We then discuss the non-linear nature of the EV charging load and simulate multiple attacks that can be launched against the power grid using these EVs. EV loads have high reactive power demand which can have a larger impact on the grid compared to residential loads. We perform simulations on two power grids and demonstrate that while the grid can recover after a 48 MW attack utilizing traditional residential loads, a smaller 30 MW EV load attack can completely destabilize the system. Finally, we suggest several patches for the existing vulnerabilities and discuss two methods aimed at detecting EV attacks.

## 1 Introduction

The power grid has witnessed a great shift, during the past years, towards automation of its monitoring and control. The resulting smart grid relies on sensors and remote-controlled actuators to perform many of the traditionally manual or localized tasks [1]. This move towards a smart grid improves grid efficiency, reduces greenhouse gas emissions, allows easier integration of renewable energy, distributed generation [2], and quicker restoration after faults. This transition, however, closely ties the grid's security to the reliability and security of the underlying smart devices and communication infrastructure [3] exposing it to cyber-attacks.

Cybersecurity has become a major concern in recent years. The most recent cyber-attack is the ransomware attack against the Colonial Pipeline in the United States that disrupted gas supplies causing a global increase in price [4]. Another major recent cyber threat is the Solarwinds hack [5] that affected around 18,000 Solarwinds clients through the compromised Orion software update.

Many of the infected users are high-security entities such as the US Department of Energy, US Department of Homeland Security, the Center for Disease Control, and multiple Fortune 500 companies. Other clients that use the infected software include NATO, the European parliament, and various governments worldwide. Other examples of cyber-attacks include the Saudi Aramco hack in 2012 when 35,000 computers were compromised threatening 10% of the world's oil supply [6]. The US OPM hack in 2014 also saw the personal information of 21.5 million users stolen [7].

Moving on to smart grids, two attacks directly come to mind, i.e., Stuxnet [8] and the Ukraine attack [9]. The 2010 Stuxnet malware attack, against Iran's nuclear facilities, represents the first major example of state-level attacks against the smart grid. A worm introduced into a Windows machine propagated to its destinations (Siemens PLC S7) and deleted itself from the untargeted devices to reduce detectability. Eventually, the malware was able to stealthily manipulate the centrifugal pressures destroying 10% of Iran's centrifuges setting their nuclear program back several years [8]. The Ukraine cyber-attack in 2015 represents the first confirmed successful cyber-attack against a power grid. The attack caused the loss of electricity for over 200,000 consumers.

Recent studies have also considered the disruption of power grid operation by compromised high wattage IoT (Internet of Things) devices such as water heaters and air conditioners [10]. While the Black IoT attack presented in [10] does not specify the IoT exploits, it is worth mentioning that attacks against IoT devices are all but inevitable. The Mirai Botnet [11] is a clear example of the weaknesses inherent to IoT devices where the attackers compromised over 600,000 devices and used them to launch DDOS attacks. IoT vulnerabilities can be summarized by the usage of weak encryption and insecure data transfer, guessable passwords, insufficient privacy protection, and a lack of secure update mechanisms.

One such emerging IoT ecosystem is the Electric Vehicle (EV) ecosystem. The EV ecosystem has incorporated multiple IoT technologies, inheriting their vulnerabilities. In fact, EVs now present a new cyber-physical attack vector [12] against the power grid. Given that transportation is a huge contributor to greenhouse gas emissions [13], there is a global push towards the usage of Electric Vehicles [14]. This, however, presents a challenge to the power utilities since improperly scheduled charging can cause high peak loads and degrade grid performance [13].

This work provides an overview of the new EV paradigm and summarizes some of the vulnerabilities that might allow adversaries to hack into the EV system and launch attacks against the power grid. We also examine the EV load profile that distinguishes it from residential loads making it ideal for an attack to disrupt grid operation. The contributions of this paper can be summarized as:

- We present an overview of the different vulnerabilities in the EV charging ecosystem as well as demonstrate multiple attack scenarios against the power grid by utilizing these vulnerabilities.

- We discuss the non-linear nature of the EV load and demonstrate how these attacks can be more harmful than traditional residential and IoT loads. We then present an attack formulation method that takes advantage of the grid conditions to maximize the attack impact through the least number of compromised EVs.
- Finally, we suggest mitigation measures to address the vulnerabilities discussed in this paper as well as 2 detection mechanisms tailored to detect EV attacks against the power grid.

The rest of the paper is organized as follows. Section 2 provides a background of the current state of EV proliferation and the impact on the grid. Section 3 demonstrates the vulnerabilities in the EV ecosystem, the properties of the EV charging load and provides an attack strategy based on stability metrics to maximize attack impact. Section 4 presents EV attack preparation and simulations as well as demonstrates the difference between attacking the grid with EV loads and residential loads. Section 5 provides suggestions for patching the presented vulnerabilities as well as two detection mechanisms and Section 6 concludes the paper.

## 2   Background

EVs still have a long path to go before they take over the car market. Range anxiety and unavailability of sufficient charging infrastructure in some areas have led to reluctance in the acceptance of EVs by some users. To this end, governments around the world offer incentives for EV purchases ranging from rebates on purchases to road tax exemptions. The public's opinion has also improved due to the decrease in prices, which can be attributed to the advancement of underlying technologies and larger production volumes. One major factor is the massive reduction in battery cost which contributes roughly to 25% of the EV price.

By the end of 2019, there were 7.3 million Electric Vehicle Charging Stations (EVCS) around the world [14] and 7.2 million EVs on the road [15]. Despite the COVID-19 pandemic, 2020 was a record sales year of 3.2 million sold EVs with 1.33 million EVs sold in China alone [16]. The Canadian EV market has also grown in 2020 with 47,000 EVs sold in 2020 [16] and a 15% increase in charging stations to reach 13,230 EV chargers at 6,016 public stations [17] [18]. Most notably, the Norwegian market has also grown in 2020 to reach 70% of new car sales at 108,000 EVs [16] and is expected to reach all EV sales in 2025 [19] earning Norway the title of the EV capital of the world [20].

The International Energy Agency (IEA) anticipates that 30 million EVs will be on the road in 2030 [14], thus reducing the demand for oil products by 2.5mb/d (or 127 million tons a year) and displacing the total emissions by 189.1 Mt $CO_2$. In line with this vision, the Quebec Government has set a target of 1 million EVs by 2030 [21] and will ban new gasoline-powered cars sales in 2035 [22]. Multiple nongovernment initiatives have also been undertaken to support the proliferation of EVs. One such initiative is the Zero Emission Transportation Association (ZETA). Two of the chief members of ZETA are Uber and Tesla that are pushing towards increased EV proliferation. Indeed, Uber's CEO expects that Uber's service will become fully based on EVs in the US and Canada by 2030 [23] [24]. Given the push towards EVs, it is important to examine the EV ecosystem components, as well as the threat EVs, introduce into the power grid. The following section presents a brief discussion of the components of the EV ecosystem.

## 2.1 Electric Vehicle Ecosystem

The EV charging ecosystem is a complex cyber-physical system composed of interconnected hardware elements, software elements and communication protocols making it an IoT paradigm at the heart of the smart grid. These elements are discussed below:

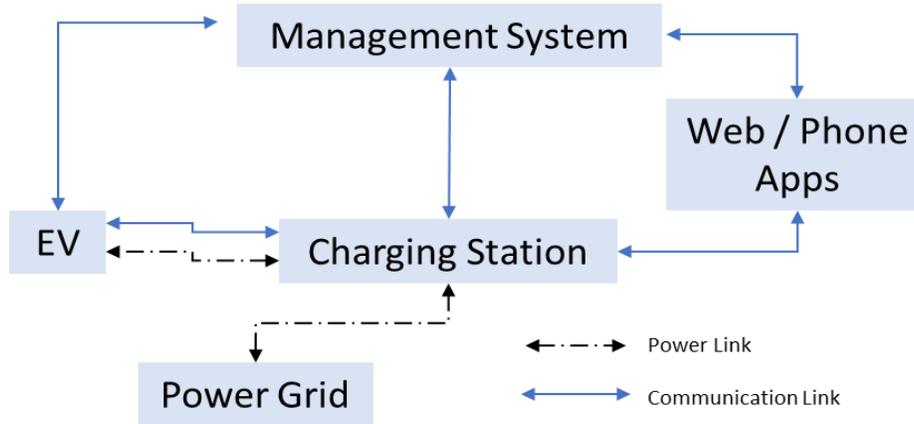

*Figure 1: Simplified Depiction of the Communication Links Between the Different Components*

- Electric Vehicles: The center element of this ecosystem is the EV itself. EVs have large batteries that require dedicated charging stations.

- Electric Vehicle Charging Stations: EVCSs are the connection point between the power grid and the EVs. The EVCS itself is an IoT device running its firmware. These charging stations are classified into 3 levels based on their charging rate [25] [26] [27]. Level 1 chargers are the slowest chargers providing a rate of 1.4 kW and can be plugged into a 110 V wall outlet but are currently being phased out in favor of the faster Level 2 chargers. Level 2 chargers require extra hardware to be installed and provide a charging rate of 7.2 kW to 11 kW. The highest level 2 rate at the time of writing this paper is 19 kW. Finally, Level 3 chargers are DC fast chargers that deliver a charging rate of 40 kW to 240 kW.

- Electric Vehicle Management System: This software is hosted on a cloud server and manages all operations of the public EVCS. Through this system users are directed to available EVCS, charging sessions are scheduled and managed and EVCS utilization data is logged. The EV management system can send the EVCS specific control signals related to the duration of the charging session, charging rate, beginning and termination commands, etc.

- User applications: These are either web applications or smartphone applications through which users can talk to the management system over the internet (in case of public EVCS) or directly to the charger over a LAN (in case of private EVCS). These services allow users to reserve and control charging sessions, pay for public charging, control charging rate, start/terminate charging sessions and monitor the status of the EV.

- Electric Vehicle Protocols: As mentioned earlier in this section, bilateral communication occurs between the components of the EV ecosystem. Figure 1 provides a simplified depiction of the established communication channels. Thus, the communication between user apps, EVs, EVCS, and the EV management system needs to be secure to ensure safe and reliable system operation. The adopted protocols vary between different equipment manufacturers, countries, and EVCS operators [28] [29]. This lack of standardization exposes the system to a variety of weaknesses as explained by the US Department of Transport [28] and summarized in Section 3.
- Power Grid: The power grid is the backbone of most modern activities including EV charging. EVCSs are connected to the power grid and draw the needed electricity from this grid and as such have a great impact on the security and stability of this grid. The following sub-section presents some of the work done on the impact of EV on the users and the power grid.

**2.2  EV charging Impact on the Grid**

As EV numbers increase further, irregular charging behavior becomes more problematic and can negatively impact the power system operation. The work performed on [30] demonstrated how uncontrolled EV charging especially during peak load time can cause loss of load up to 6.89%. In fact, it is estimated that a 10% EV penetration in Portugal can cause significant voltage-drop at peak times [31]. The work performed in [32] performed a comparison between two optimization objectives, to reduce the peak load by shifting the EV charging to night times and by utilizing the Vehicle to Grid reverse power flow to reduce day peaks respectively. Their work demonstrated that reducing peak load and minimizing operation costs cannot be achieved simultaneously. The authors concluded that it is more important to properly manage the EV charging schedule rather than EV discharging to improve system load without significantly increasing the operation cost.

Another aspect of EV impact on power system cost was studied in [33] that focused on power system infrastructure investment cost as well as system losses. This study performed simulations on a residential Area A with 6,000 consumers having 3,676 cars and a mixed residential/industrial Area B having 61,000 consumers and 28,626 cars. Multiple EV penetration levels were considered between 35% and 62% of the total car numbers. The EV charging stations were spread randomly in the grid at locations that already had non-EV loads. The simulations were performed for the peak and off-peak demand scenarios at each area and the results show increased investment cost and system losses at all levels of EV penetration. The incremental investment was found to be 201/EV in Area B and 6310/EV in Area A since it was already operating close to its operating limits. The authors also simulated the impact of the EV load on system losses and concluded that the losses could increase by up to 30-40% under a 62% EV penetration scenario.

**2.3  EV Attacks Against Users and the Grid**

Several works in the literature have discussed attacks against or through the EV ecosystem against the users and the power grid. Attackers that gain control of the EV Battery management system through compromised web services or malware downloaded to the EVs systems can cause severe damage to the EV itself. In fact, [34] and [35] discuss how the attackers can cause damage to the EV batteries by manipulating the charging current and bypassing the safety measures in place. However, attacks against the users are considered out of the scope of this work.

The authors of [12] presented an EV attack formulation by relying only on publicly available data to destabilize the Manhattan power grid. Their method consists of representing the power system as a feedback control system and the EV as the feedback gain of the system to determine the number of EVs required. Their work concluded that although Manhattan doesn't currently have enough EVs to mount such an attack, the growth in the EV numbers will soon provide a large enough surface to make it possible. Their work, however, relied on the DC power flow model that completely ignores the reactive power flow and other grid behavior issues. In contrast, our work will take into consideration the nature of the EV load presented in Section 3.5 that makes it ideal to mount attacks against the power grid with a larger impact than pure active power loads and loads with a high power factor.

## 3    EV Vulnerabilities, Load Properties and Attack Strategy

Being IoT devices that adopt multiple communication platforms and web applications, the EV infrastructure inherits the vulnerabilities of the adopted technologies [36]. The pressure to achieve rapid expansion has also been considered a great hindrance to the secure deployment of the EV infrastructure. Operators and manufacturers often forgo security measures to achieve faster and cheaper deployment of their equipment. Due to the connection of the EV infrastructure to the power grid, the secure operation of EV charging is considered pivotal for the security of the modern smart grid. To this end, we present the following insight into the security issues surrounding the EV charging ecosystem. The US Department of Transport mentions the following four vulnerabilities in the EV ecosystem [28]:

- EV Charging Infrastructure Lacks Cybersecurity Best Practices: The EV industry lacks secure software design and development methodologies.
- EV Charging Infrastructure Lacks a Trust Model: There is no agreement on a secure communication standard for communication between EVs and the EVSCs.
- EV/Charging Infrastructure Lack Cybersecurity Testing: There are inadequate integrity protections and there are limited cybersecurity monitoring tools to detect malicious activity.
- Commercial Charging Infrastructure Lack Physical Security: Most EVCS are exposed and can be physically accessed and tampered with. Different EVCSs have different physical features.

Given the mentioned vulnerabilities, this section discusses possible EV ecosystem exploits and specific system vulnerabilities. Then we discuss the EV charging load properties that make it ideal for attacking the grid as well as present an attack strategy.

## 3.1 EV Ecosystem Exploits

The National Institute of Standard and Technology (NIST) [37] classifies attacks against the EV infrastructure as:

- Physical: EVCSs lack physical security and malicious actors can damage the EVCS, steal electricity, or even install malware through the available USB ports. EVCS can even be blocked at public locations by parking Internal Combustion Engine vehicles in front of them (known as ICEing) to cause a denial of service.

- Local: This exploit is based on gaining logical access to the EVCS through the firmware vulnerabilities. The firmware updates of some vendors such as Schneider Electric are available online and can be disassembled, and reverse engineered by attackers to find their weaknesses and possible entry points into the system [28] [38]. Kaspersky labs were able to reverse engineer the firmware of ChargePoint home chargers [39].

- Limited Remote: Attackers can take advantage of Local Area Networks at homes or public charging locations to access the EVCS. These networks usually employ weak credentials and outdated encryption techniques. The communication between the EV and the EVCS takes place through a set of protocols over the charging line adding vulnerabilities to the system.

- Remote: As mentioned above, EV users communicate with the EV management system online. Whether this communication takes place through a webpage or a mobile application, it creates a plethora of entry points for malicious actors. Furthermore, the EVCSs communicate with the EV management system through protocols such as the Open Charge Point Protocol (OCPP).

## 3.2 EV to EVCS Communication Vulnerabilities

The IEC and ISO protocols define the communication between the EV and the EVCS. Information is exchanged over the powerline through a dedicated control pin. However, this communication has its issues and vulnerabilities as explained herein.

### 3.2.1 IEC Vulnerabilities [40] [41]

IEC protocols are well established in the Control of substations but have also been included in the EV ecosystem. These protocols define multiple aspects of EV charging, but we only mention two of them herein. IEC 61850-90-8 fulfills smart charging requirements and has considered other standardization efforts from the beginning. However, basic functionality for EV charging like user authentication was considered outside the scope of this protocol and was delegated to other protocols like OCPP or other IEC protocols. IEC 61851-1 defines a safety-related signaling mechanism between EVs and EVCS based on Pulse-Width Modulation (PWM). The PWM signal is transferred over the control pilot pin of the charging ca as demonstrated in [42].

### 3.2.2 ISO 15118 Vulnerabilities

The ISO/IEC 15118 defines an international complementary standard to IEC 61851-1 and provides bidirectional digital communication [41]. However, ISO 15118 is not demand response compliant and does not devote a space for privacy except by stating that private information shall only be transferred, when necessary, to the intended addressees [43]. External studies have highlighted the potential privacy concerns related to the usage of this protocol [43] and the work done in [42] even performed a real-world attack campaign. The authors of [41] [42] and [43] highlighted some weaknesses in the protocol or the improper use of the available security measures.

- Signal-Level Attenuation Characterization (SLAC) is a protocol that can operate in a secure mode, with mutual authentication and encrypted communication.
- TLS encryption is supported by this protocol but is dropped when the charging session is authorized by an external source such as RFID (Radio-frequency identification) cards, mobile app networks, or manually by an operator.
- Public key infrastructure can also be implemented with ISO 15118.

These security measures, however, are optional and have been mostly ignored by manufacturers and operators to reduce additional cost and overhead leaving the communication through plain text vulnerable to attacks.

**3.3    EVCS to Management System Communication Vulnerabilities [40] [44] [45]**

OCPP coordinates communication and power flow between EVCS, EV management system, EVs, and the grid and allows full remote control of the EVCS in real-time. OCPP is leading the efforts towards standardization and has become a De Facto open-source protocol and is currently the most widespread charging protocol. OCPP supports online change in charging configuration, as well as starting and ending charging sessions and online billing. OCPP supports smart charging by controlling variables such as session scheduling, charging rate, and duration of charging. OCPP communication uses simple HTTP to manage EV charging and adopts diverse communication infrastructures. Although OCPP 2.0 and 2.0.1 were released in 2018 and 2020, OCPP 1.5 and 1.6 remain the dominant versions deployed. OCPP 1.6 adopts diverse communication infrastructures such as SOAP/XML (extensible markup language) or JSON (javascript object notation) over WebSockets (WS). OCPP has an optional TLS layer for secure communication but has been mostly ignored by manufacturers and operators to reduce overhead and price. OCPP 1.5 had an optional hash function that can be used for extra security. However, instead of making it mandatory in newer versions, OCPP 1.6 eliminated this option. The main security concern for OCPP is ensuring that a charging session is authorized by a billing system. This along with communication using plain text and the absence of wide adoption of encryption allows attackers to highjack the communication and gain control of EV charging. Attackers that are unable to break the encryption, can capture messages such as those that start and terminate charging sessions and use them in replay attacks to disrupt the EV charging. OCPP 1.5 and 1.6 also offer a functionality known as local authorization list (LAL), which allows an EVCS to serve customers even if it gets disconnected from the Management System. An attacker, however, can use this functionality to force the Management System into accepting EV charging requests without proper authentication.

### 3.4 Firmware / IoT Vulnerabilities

As mentioned in Section 2, the EVCS host a firmware the controls their operation and interacts with multiple online services available for management and user interaction with the EVCSs. As IoT devices, EVCS exhibit the same types of vulnerabilities present in other IoT devices. In addition, the web pages and phone apps used in this environment also exhibit the same vulnerabilities as other web/phone apps. These vulnerabilities are discussed in the OWASP Top 10 Application Security Risks [46] and can be discovered through the framework presented in [47]. Other vulnerabilities include malicious update packages sent to replace the authentic updates by the manufacturers similar to the Solar Winds Hack [5] that would allow the attacker to embed their code in the deployed system and virtually control all aspects of the charging process. The most critical vulnerabilities that were proven to be present in the EV ecosystem in [28] [38] [39] and by work involving one of the authors of this paper are summarized below:

- SQL Injection: allows the attacker to gain access to privileged user information and manipulate the EVCS firmware.
- XML/External Entity Injection: allows the attacker to inject HTTP requests into the system and in some cases gain remote access to the EVCS.
- Server-Side Request Forgery (SSRF): allows the attacker to redirect traffic towards internal/external endpoints causing denial of service and reading files and record logs of the EVCS.
- Cross-Site Scripting (XSS): allows the attacker to inject malicious code into the EVCS allowing them to highjack user accounts or even administrator accounts in some cases.
- Comma-Separated Values (CSV) injection: allows attackers to embed XSS payloads that get triggered and stored on the EVCS leading to hijacking administrator session tokens. Using this vulnerability, attackers can manipulate EVCS functionality.
- Cross-Site Request Forgery (CSRF): allows the attacker to induce target users to perform unintentional actions that lead to setting modification and manipulation of EVCS functionality. This weakness can allow attackers to gain control of the EVCS.
- Hard-Coded Credentials: are utilized by developers to ease the coding process, but it allows attackers to recover the hardcoded login credentials in the source code of the EVCS or the associated application and gain unauthorized access to the EVCS.
- Missing Authentication: allows the attacker to gain unauthorized access to user accounts without being properly verified by the EV management system.

### 3.5 Electric Vehicle Charging Properties

Unlike conventional residential loads, EVs are battery storage loads that present a new concern to the power grid. Charging this battery is a non-linear process that would reduce the quality of the delivered power [48] – [53]. The EV charging load would increase the reactive power demand and reduce the Power Factor (pf), introduce harmonics, cause voltage distortion and fluctuation, and it can also cause three-phase imbalance [48] – [53].

EV charging is a typical non-linear load based on rectifier circuits and power converters [49]. Non-linear loads draw non-sinusoidal current from a sinusoidal voltage [50] causing voltage and current waveforms to have different shapes. Some mainstream EVCS utilize uncontrollable rectifier circuits having high harmonic distortion. Other EVCS that support V2G technology, utilize controllable power converters [51]. These non-linear loads contribute to a poor power quality on the grid [52]. New solid-state devices are feasible using MOSFETs and IGBTs as well as incorporating active and passive filters into the chargers [52]. Yet these devices are costly, bulky and incur losses that reduce the overall efficiency. Regardless of the specific circuit of the EVCS, adding a large amount of non-linear load would increase the reactive power demand of the system [53]. As the work in [51] demonstrated, nonlinear loads can have a power factor as low as pf=0.15 lagging. The authors also discussed that although power factor correctors can improve the pf, a perfect pf=1 is not possible due to the harmonic distortion of these non-linear loads. The authors of [48] also demonstrated that an AC-DC converter (rectifier) can have a power factor as low as pf=0.4.

Uncompensated battery chargers have a power factor slightly above or below 0.6 [53]. To make things worse, an attacker that gains control of an EVCS with controllable power converters can manipulate it to draw more reactive power from the grid [50]. As a middle ground, we consider the EV attack loads will have a pf=0.6 lagging [54].

Transmitting large amounts of reactive power from generators to the loads causes large transmission losses in the power lines, reduced voltage levels, and might cause severe problems in terms of grid stability. Synchronous generators (not being operated as condensers) are limited in terms of the total reactive power they can produce at a given point of active power generation [55].

On the other hand, traditional residential loads and home IoT loads have a high-power factor being mostly resistive. This is especially true for the larger loads such as electric water heaters and space heaters that have a power factor very close to pf=1. Other large loads such as air conditioners and refrigerators have a power factor of 0.8<pf<0.9. Although recent loads such as laptop chargers and LED lights and smart TV are power electronic-based and have a nonlinear load, their size and overall contribution to the total residential load remains acceptable where the average residential load is usually considered to be higher than pf=0.8. Due to this distinct nature and size of the EV loads, their impact will be much larger than that of the residential and IoT loads.

### 3.6 Smart EV Attack Vector

Attackers with knowledge of the power grid can craft smarter attacks that target the power grid with smaller numbers of compromised EVs and still disrupt the power grid operation. Attackers can gain access to insider information from the power utility by exploiting multiple vulnerabilities such as the weakness in network and communication devices, third-party services, and intrusion through the supply chain [56]. Also, there is a lack of widespread employment of cybersecurity personnel as well as the lack of cyber hygiene for the utility employees. Attackers can also introduce malware into the utility's control system through insiders (knowing or unknowing) or through software updates similar to the Stuxnet [8] and SolarWinds [5] attacks respectively. With knowledge of the power grid, attackers can use existing power grid stability methods to locate buses where their attack impact can be amplified. PV and QV curves [57] [58] are two examples of such methods. Multiple studies have considered the estimation of power grid topology by monitoring the measurements of the power grid [59]-[67]. These methods are summarized as:

- Monitoring voltage and loads at terminal buses.
- MILP programing and monitoring the response to load perturbation.
- Perturbing inverter injections and monitoring system reaction.
- Machine learning approximation of parameter or line impedances based on PMU measurements.

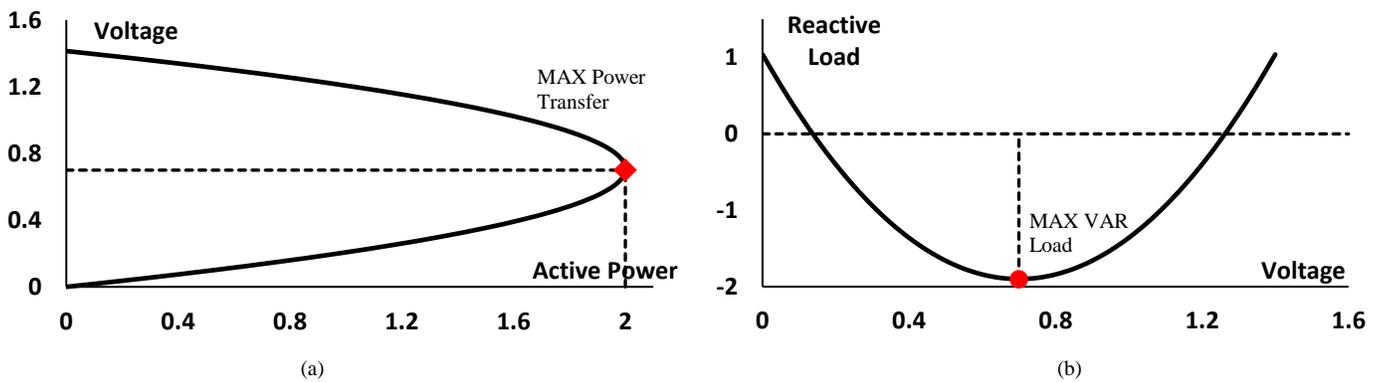

*Figure 2: (a) Typical PV Curve and (b) Typical QV Curve based on PowerWorld Manual on PV and QV Curves [73]*

### 3.6.1 PV and QV Curves

PV curves monitor the voltage at target buses due to load and generation increases in the system. The curve is generated by selecting a set of loads and increasing them incrementally. Specific generators are selected to supply the extra load. At each iteration, the power flow equations are solved, and the voltage value is recorded. A curve like Figure 2 (a) is generated where the knee point, marked with a diamond, represents the maximum extra load that can be added before voltage collapse occurs. QV curves monitor the voltage at a bus due to increased reactive power demand in the system. The curve is generated for a certain bus by adding a fictitious generator at that bus and varying the VAR output. A curve similar to Figure 2 (b) is generated where the valley point, marked with a dot, is the maximum reactive power that can be drawn before voltage collapse.

### 3.6.2 Attack Strategy

After performing grid reconnaissance, the attackers can generate the PV and QV curves for the system buses to determine the active and reactive load needed to cause the collapse. However, the attacker can use these methods to locate the weakest buses that are more likely to cause larger disturbances once attacked. By attacking these buses, the attacker can disrupt the system while simultaneously minimizing the needed number of compromised EVs to reduce the attack cost and difficulty. The PV curves and Q-V Sensitivity curves were generated for the grid in Figure 10. Indeed, buses 1, 2,3 and 4 were more sensitive to increased load than buses 5,6, and 7. In fact, attacks on buses 6 and 7 caused much smaller variations in the system behavior.

The V2G capability of the electric vehicles can also be incorporated by considering it as part of the generator vector in the PV curve. These stability methods can be performed for the system base case as well as in the presence of contingencies. Attackers that can simultaneously launch physical attacks, such as disconnecting lines, can use this option to determine their attack impact beforehand. These PV and QV curves can also be plotted to take into consideration the presence of VAR compensators and tap-changing transformers. By utilizing these features, the attacker can craft the attack to circumvent these protection methods.

## 4 Case Study: EV Attacks Against the Power Grid

Power grid stability depends on the balance between the demand load and generation supply. EVs can be used to perform the same types of attacks as residential loads to disturb this balance and new attacks not previously possible through other loads. However, the special nature of EV loads presents a surface for attacks that were previously not possible. Attacks through the EV charging load can cause larger disturbances while utilizing smaller loads. In addition to that, due to the nature of the power flow equations, frequency can sharply increase or decrease in response to sudden drops or spikes in loads. Voltage also needs to be maintained within acceptable limits, usually 5% of the nominal value and is tightly connected to the reactive power flow. Due to the nature of generators and power systems, reactive power is more difficult to generate and push onto the power grid than active power [53]. Due to the nonlinearity of battery charging, an EV load's high reactive power demand would have a larger impact on the voltage.

When voltage and frequency violations exceed a given limit, utilities take corrective action to restore the normal levels to avoid damage to the devices connected to the grid. Persistent violations can cause the operator to start load shedding or might eventually force the grid to go into a state of blackout. In the following section, we discuss the possibility of the EV attacks and present general EV attack scenarios as well as others specific to electric vehicles. We also compare the impact of the EV load to the impact of traditional residential loads when used to attack the grid. We then perform simulations to demonstrate the attack results.

### 4.1 Attack Preparation

Being an integral part of the power grid, the security of the EV ecosystem seriously impacts the secure operation of the grid. By exploiting the EV ecosystem's vulnerabilities, attackers can manipulate EV charging to disrupt the grid [68]. The following section discusses how an attacker can gain access to the EVCSs and the available load at the attacker's disposal through this ecosystem.

#### 4.1.1 EV Locations and Load Size

Attackers can locate EVCS using publicly available maps provided by the EVCS operators or by third parties. Chargepoint for example offers maps that contain the locations of EVCS offered by them and other manufacturers [69]. Other maps are offered on websites such as Plug share, Chargemap [69], and the Alternative Fuels Data Center (AFDC) [70]. Using this data, attackers can know the location, power rating and historical utilization profiles of EVCS to plan their attack.

To demonstrate the number of EVs and the EV load in comparison to the total system load we present the following discussion. We acknowledge that the current numbers of EVs might not be sufficient to cause the intended damage, but the following example demonstrates its future possibility. For instance, the total load for the island of Manhattan is between 2,000 MW and 2,100 MW [12] and the total number of cars is 2 million [71]. Table 1 represents the total EV load at different EV penetration levels between 10% and 50%. The table also demonstrates the total load at the different common Level 2 chargers and the available fast chargers.

*Table 1: EV Load in MW as a Function of EV Penetration Levels and Charging Rate*

| EV Penetration Levels | Level 2 Chargers | | | Level 3 Fast Chargers | |
|---|---|---|---|---|---|
| | 7.2 kW | 11kW | 19 kW | 40 kW | 240 kW |
| **10%** | 1,440 | 2,200 | 3,800 | 8,000 | 48,000 |
| **20%** | 2,880 | 4,400 | 7,600 | 16,000 | 96,000 |
| **30%** | 4,320 | 6,600 | 11,400 | 24,000 | 144,000 |
| **40%** | 5,760 | 8,800 | 15,200 | 32,000 | 192,000 |
| **50%** | 7,200 | 11,000 | 19,000 | 40,000 | 240,000 |

As demonstrated by Table 1 the total EV load can surpass the total load of the system. In our attacks in the sections that follow, the EV load varies between 3% and 6% of the total system load and only in one of the cases do we utilize 19% of the total system load in Section 4.4.2. Assuming the case of 50% EV penetration and a charging rate of 7.2 kW, the attacker would only need to compromise 5% of the available EVs (2.5% of the total number of cars) to achieve the 19% compromised load in Section 4.4.2. We assume a charging rate of 7.2 kW to offer a restrained example of the EV load. In reality, the 7.2 kW rate is disappearing in favor of the newer 11 kW rate that is commercially available. This example serves to demonstrate that with the global push towards EVs, the attack magnitudes are easily achievable by an attacker aimed at disrupting grid operation.

### 4.1.2   EV Botnet Command and Control

Attacks that exploit the EV ecosystem vulnerabilities can be summarized as:

- Passive eavesdropping [42] [68] during which the attacker collects data by reading the exchanged messages.
- Jamming attacks during which an attacker can delay or even drop messages to disrupt the communication between the Management System and the EVCS.
- Denial of service attacks can also be launched making certain stations unavailable to users (or appear unavailable) to redirect EV traffic or disrupt charging.
- Tampering with scheduled charging times and charging prices to redirect EV traffic.

- Fully compromising the communication channel to achieve a Man in the Middle and fabricate or modify any message being sent from and to the EVCS. This would grant an attacker great freedom to do whatever they want.

Through the vulnerabilities mentioned in Section 3.4 attackers can gain control of the EVCS and create a botnet of compromised chargers that can be used to target the power grid. Through attacks such as XSS and CSV attackers can gain access to user accounts and even admin accounts on the EVCS. The attacker can also gain access to some accounts by utilizing the hard-coded credentials and missing authentication. By controlling the user accounts, the attacker can start/end charging sessions as well as change the charging rate and any functionality available to the users. By controlling the admin accounts, attackers can initiate any functionality that is supported by the EVCS but not made readily available to the users. An example of this is the variable charging rate of the deployed EVCS that might not always be available for users but is possible using the deployed hardware. By launching more serious attacks such as SQL and XML injections, the attacker can gain full control of the EVCS and virtually perform any action that is possible on the deployed hardware. Using the discussed vulnerabilities, the controlled botnet of EVCS can be used to launch multiple attacks on the power grid by synchronizing mass charging and discharging sessions to disrupt the power grid operation.

## 4.2 Simulation Setup

To demonstrate the impact of the attacks, we perform simulations on the WSCC 9-bus system shown in Figure 3 and the 7-bus test case taken from the Glover et all textbook [72] shown in Figure 4 that are commonly used for academic research. To this end, we use PowerWorld [73] which is a power system simulator that is used to perform power flows, contingency analysis, transient stability, optimal power flow, and other power system operations. For this study, we use the free version of PowerWorld that can accommodate up to 13 buses. As discussed in Section 3.5, we consider the power factor of the EV charging as pf=0.6 lagging [53].

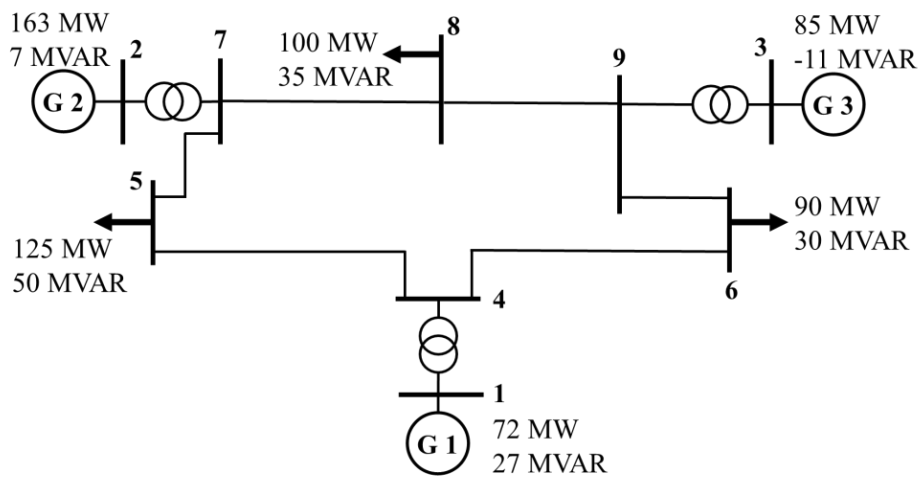

*Figure 3: WSCC 9-Bus System*

Given the dependence of the grid's transient behavior on the generator/turbine models, we used the control models common to similar studies. On a special note, the attacks achieve similar impacts under different controller setups. The general shapes of the curves remain the same, but the exact values might differ. However, this demonstrates that the attacks can be successful under different conditions, but the magnitude might be scaled slightly up or down to achieve the desired impact. The used models are:

- Machine Model: GENSAL
- Generator Exciter Model: IEEE T1
- Turbine Speed Governor: IEEE G2
- EV V2G injection: Negative fixed MW Load to represent a controlled power injection.

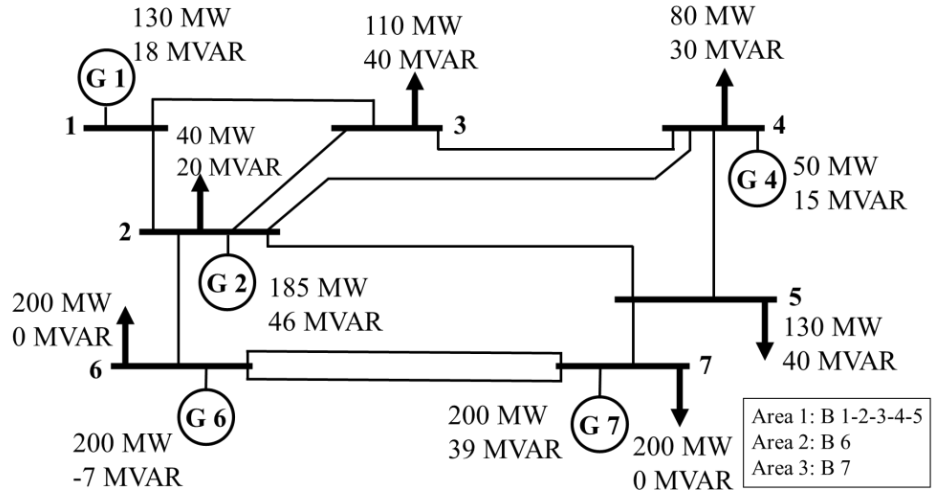

*Figure 4: Glover 7-Bus System*

### 4.3 EV Load Manipulation Attack Scenarios

EVs can be used to launch the same types of attacks as any residential load such as manipulating the power demand to cause line trips and cascading failure. This is demonstrated by attacking the 7-bus system. Through the controlled EVCS mass synchronized charging can be initiated to increase the demand at bus 3 by 50 MW of EV load (82 MVA). This would cause the line connecting bus 1 to bus 3 to overload and trip. After this initial trip, the line connecting bus 2 to bus 3 would be overloaded and would trip. Subsequently, the lines L3-4, L2-4 and L2-5 would be overloaded leading to multiple failures followed by a blackout.

Another attack that takes advantage of load manipulation is creating a surge in demand to cause a frequency drop on the power grid. Such an attack is simulated against the 9-bus system by adding 7.2 MW of load representing 1000 level 2 charging EVs at each of the 3 load buses. The attack was launched at t=15s and the frequency response is presented in Figure 5. Due to the increased power load, the generators would slow down resulting in a frequency drop on the power grid. When the frequency drops below the preset threshold, the generators are disconnected from the grid eventually leading to further reduction in frequency and possibly a blackout. This attack can be made worse by initiating it during peak times when the grid has fewer reserves to supply the extra demand. Also, when the renewable energy resource (RES) share is high, the attack can have a larger due to lower grid inertia.

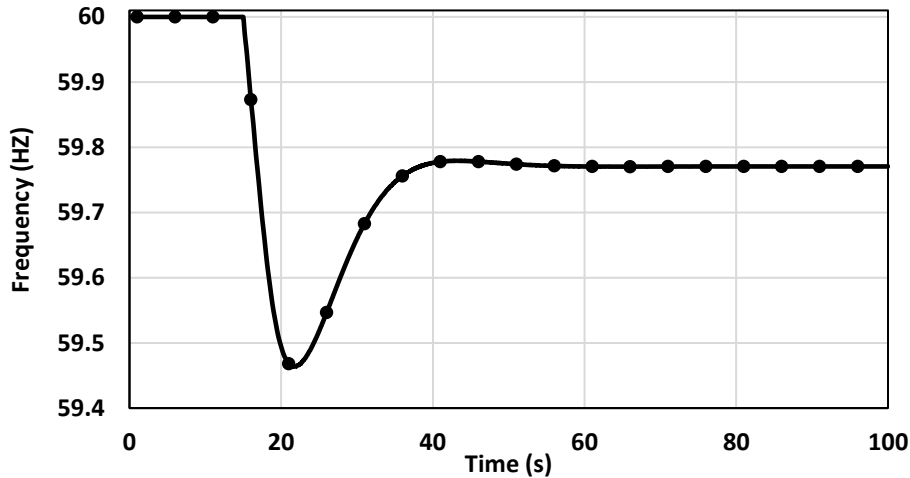
*Figure 2: Frequency Drop on 9-Bus System Due to Increased Demand 3000 EVs*

**4.4 EV Specific Attack Scenarios**

The nature of EVs allows for new types of attacks not available through traditional residential loads. These attacks are made possible through the reverse power flow feature that allows EVs to inject the energy stored in their batteries into the grid. This feature, however, can be maliciously exploited by attackers to create injections that can disrupt the grid operation. In fact, attackers can inject power at a power factor controlled by them to maximize the attack impact causing disruptions larger than reduced load. The power injection is even more impactful than adding a generator because the injection does not participate in generation control and voltage control. This attack would also greatly impact the grid's power factor, giving this injection a larger impact on the voltage and frequency. Two attacks that utilize this feature are presented in the subsections below.

**4.4.1   Controlled Vehicle to Grid Injection (V2G) Attack**

To demonstrate this attack, we added a negative load of -50 MW and -10 MVAR at bus 5. A negative load would resemble a power injection by a botnet of compromised EVs forced to discharge at the same time. As discussed earlier, this sudden injection would disrupt the balance between the total load and total generation causing frequency and voltage violations. The attack was performed by connecting the power injection at t=15s which caused a frequency rise and voltage violations at multiple buses. Due to the imbalance between the extra generation and the load, the system frequency rises to 61.6, and the generators' rotational speed would increase causing the frequency behavior represented in Figure 6.

Similarly, Figure 7 represents the voltages at buses 4, 5, 7 and 9 that increase due to the added generation. The voltage at all buses experienced an increase above 1.05 pu during the transient response, but these graphs represent the voltage at the buses that remained greater than 1.05 after the system went back to steady-state. The same experiment was performed by replacing the power injection with a generator to compare the impact of a controlled injection to a regular generator. This alternative attack resulted in smaller violations and required 51 MW to reach the frequency limit of 61.5 Hz and caused no voltage limit violations. This demonstrates that a smaller controlled V2G power injection has a higher impact than a rotating generator whose control system can react to the grid conditions.

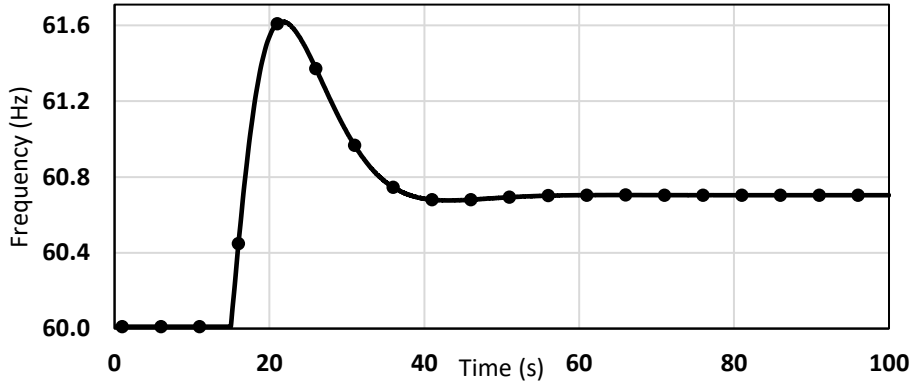

*Figure 3: Frequency Rise on 9-Bus System Due to Power Injection*

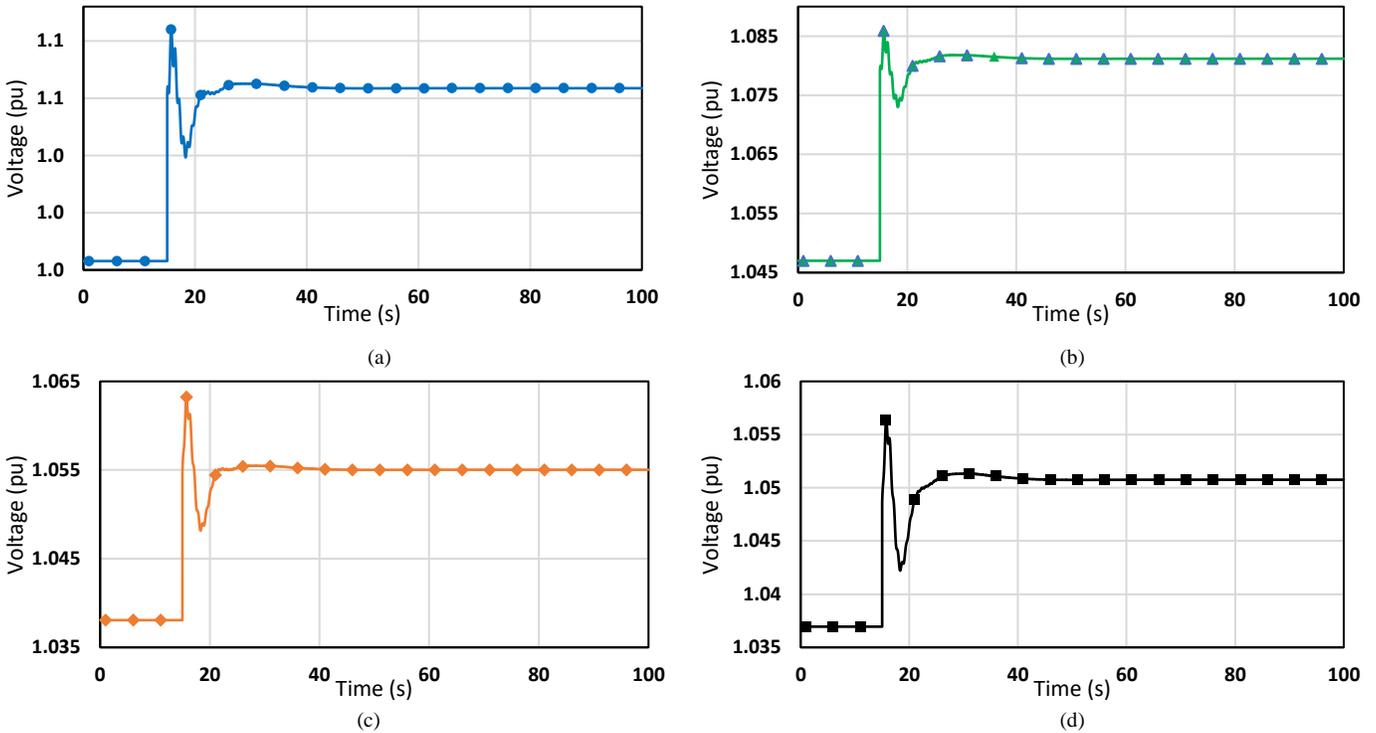

*Figure 4: Voltage Limit Violation at (a) Bus 4 (●), (b) 5 (▲), (c) 7 (♦) and (d) 9 (■) Due to Power Injection*

### 4.4.2 Switching Attacks

The second type of attack is a switching attack, which is performed by alternating between increased power injection and increased power demand. This attack is initiated by increasing the power demand to cause a frequency drop as in the attack in Section 4.3. The second step of this attack happens when the system starts its recovery. The attacker would switch off the EVCS initiated in the first step and initiate a power injection to cause a frequency increase that is amplified by the operator's effort to increase the speed in response to the extra load. The attacker would then alternate between these steps for the desired duration.

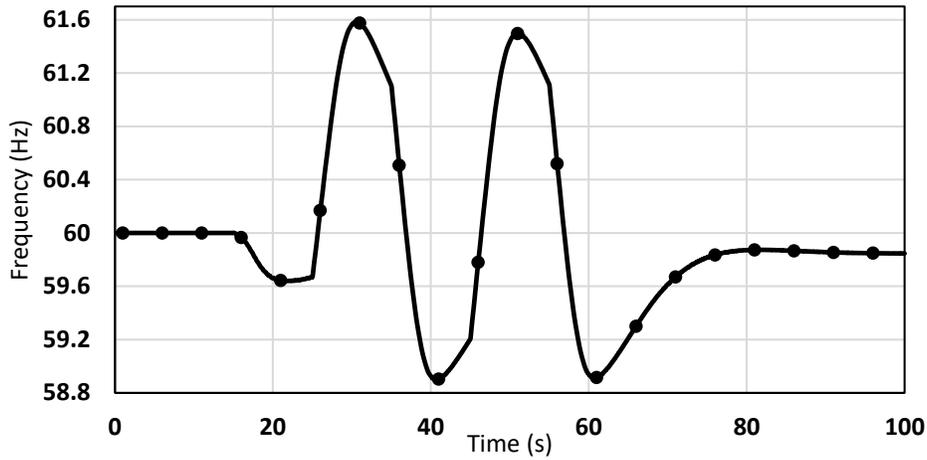
*Figure 5: Transient Frequency Response to the Switching Attack*

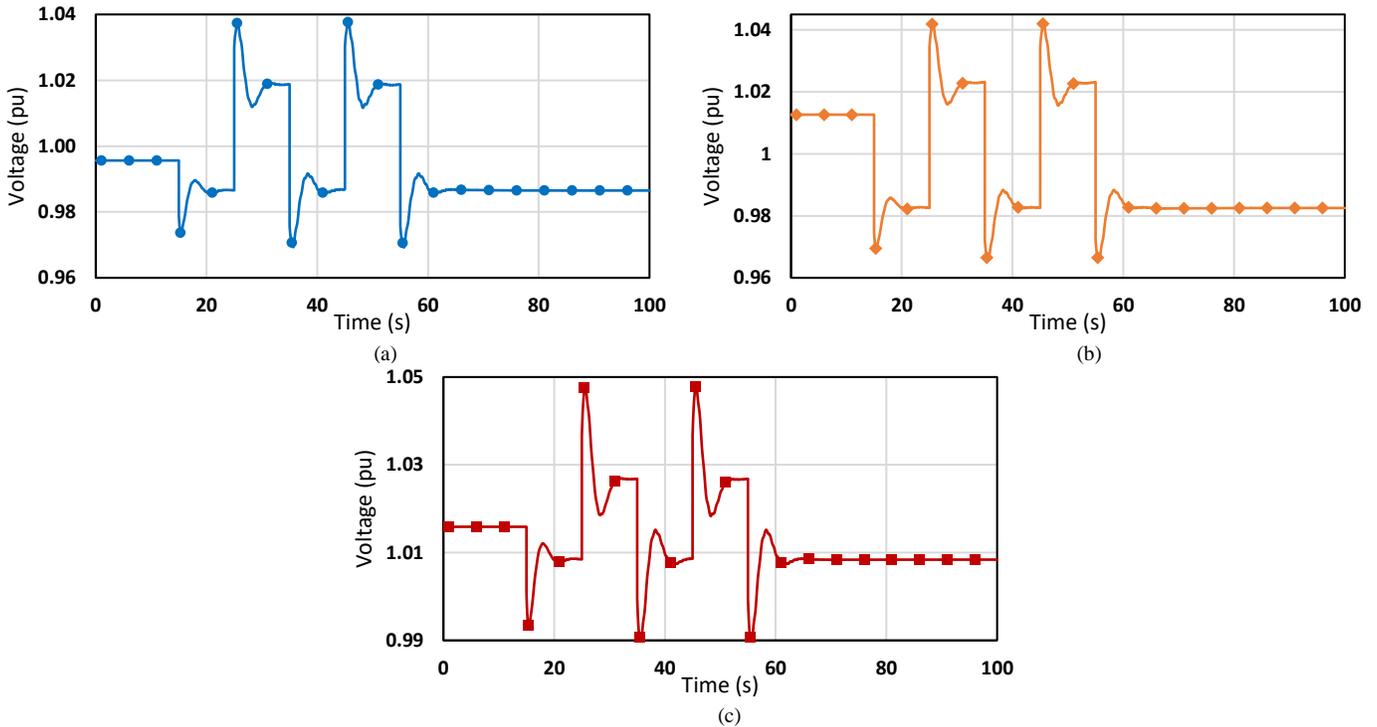
*Figure 6: Transient Voltage Response at Load (a) Bus 5 (●), (b) Bus 6 (♦) and (c) Bus 8 (■) Due to the Switching Attack*

We simulate this attack against the 9-bus system by adding a 21 MW (highly inductive) load at bus 6 representing 2900 EVs charging at 7.2 kW and a negative 60 MW load (power injection) at bus 5 representing 8300 discharging EVs. The attack is initiated by closing the load at bus 6 at t=15s then opening it and closing the V2G injection at bus 5 at t=25s. This process is repeated at t=35s, 45s, and 55s. This attack causes the frequency to fluctuate as depicted in Figure 8 as well as frequency limit violations while simultaneously causing the voltage fluctuations at the load buses shown in Figure 9. The importance of this attack is that it does not require huge loads or injections to cause behavior depicted in Figure 8 and Figure 9. Even when the compromised EV numbers are much less than the example above, a sustained switching attack can cause the frequency and voltage oscillations hindering the system's return to normal operation. A sustained load oscillation would damage the turbines due to the constant acceleration and deceleration. Other equipment such as transformers might require more frequent maintenance intervals if the attacks are sustained for long durations. The performance of the appliances connected by the consumers might be impacted as well but this is beyond the scope of this paper.

It is important to mention that when the power injection increase is accompanied by a demand drop at bus, 5 the impact is significantly amplified. The normal demand at bus 5 is 125 MW. When this demand is decreased by 20% simultaneously with the power injection, the frequency response goes over 62.5 Hz. This would trigger multiple protection mechanisms further exasperating the problem depending on the grid protection mechanisms and the regulations (varies by geographic location). A frequency increase of more than 4% will trip the generators and cause a blackout.

### 4.5 Other Attack Implications

Even when the controlled botnet of compromised EVs is not large enough to cause frequency and voltage violations, the attacker can still launch attacks that disrupt power grid operation. By increasing the demand even by moderate amounts, an attacker can shift the system away from the optimal dispatch point and cause extra costs to be incurred on the grid. EV charging is usually scheduled during low-demand periods to benefit from the low electricity cost. However, when an attacker forces the compromised EVs and EVCSs to charge during peak times, the increased load on an already taxed grid would cause increased transmission losses, voltage imbalance, and the use of peak load generators. The latter are generators available to the utility to supply loads when the baseload generation units cannot serve the entire demand. Peak generation units are generally more expensive to operate but are used due to their flexibility and ease to ramp up the generation and short startup time. Also, the increased cycling of these units between on and off would require shorter maintenance intervals and increased maintenance costs.

Sudden spikes in demand or generation also cause oscillations of the generator output. The increased varying in generator output can also have an impact on the lifetime and maintenance frequency of the generators. The increased load, especially the reactive nonlinear EV load, would add Time Distortion Harmonics into the system. EV attacks can also theoretically be used to hide transmission line outages similar to False Data Injection attacks [74]. By manipulating the load at both ends of the disconnected line the attacker might be able to hide the line outage from state estimators used by the grid operator.

### 4.6 Comparison to non-EV Attacks

In this sub-section, we aim to demonstrate how attacks launched through EV loads can be more impactful than those launched through traditional residential loads. For this purpose, we will simulate our results on a modified version of the 7-bus system presented in Figure 10 also taken from Glover et al [72]. The total load on this system is 800 MW and the transmission line limits have been increased. The Optimal Power Flow has been solved on PowerWorld and the extra load to be supplied will be divided among the generators based on the cost of the extra generation.

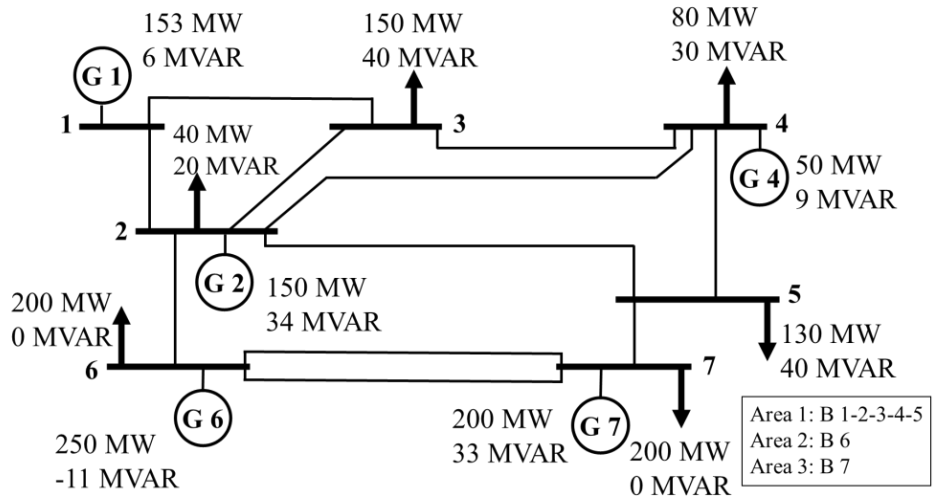

Figure 7: Modified 7-Bus System

### 4.6.1 Residential/Home IoT Load Attack

We will first perform an attack using traditional residential loads having pf=0.8 as described in Section 3.5. This attack was simulated by adding a load of 48 MW (60 MVA) divided equally on each of buses 2, 3, 4 and 5 (12 MW each). The sudden load increase at t=15s causes a frequency drop as demonstrated in Figure 11. Although the frequency dropped to 59 Hz, the system was able to maintain the transient stability as all the areas, and generators remained in synchronism and the values returned to a steady value. As for the steady-state stability, the system did not exhibit any voltage violations and all parameters remained within acceptable limits. After the system stabilizes, the system operator could ramp up generation beyond the automatic control to go back to the nominal frequency. This is to demonstrate that although the frequency dropped to 59 Hz which might initiate load shedding, the system was not completely destabilized as the case is for the attacks in the following subsections.

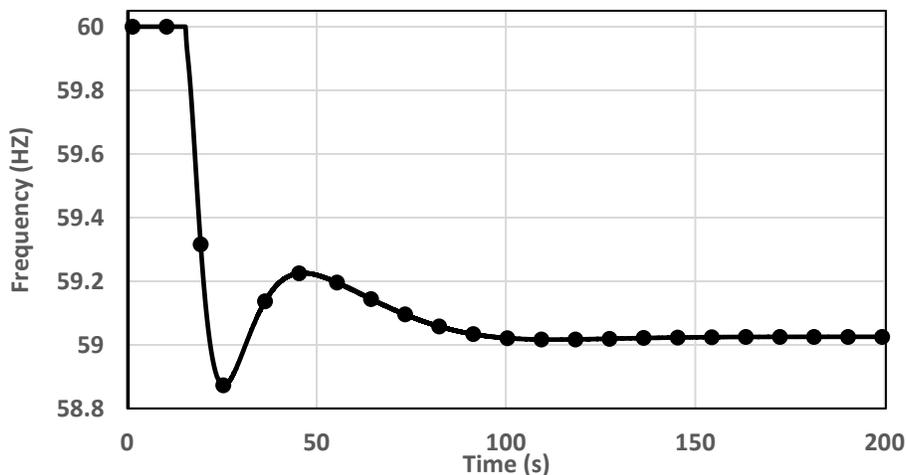

Figure 8: 7-Bus System: Transient Frequency Response After the 48 MW Attack

### 4.6.2 EV Load Attack – Similar MW load (6000 EVs)

The first EV attack is launched against the same locations as the attack in Section 4.6.1 (bus 2, 3, 4 and 5). This attack was performed with an EV load having the same active power demand of 48 MW. At the pf=0.6 assumed for this type of load, the apparent power is equal to 80 MVA instead of the 60 MVA of the previous attack. This attack represents a synchronized mass charging of 6000 EVs charging at a rate of 7.2 kW. The EVs are split equally among the 4 buses leading to the 4 equal loads of 12 MW (20 MVA).

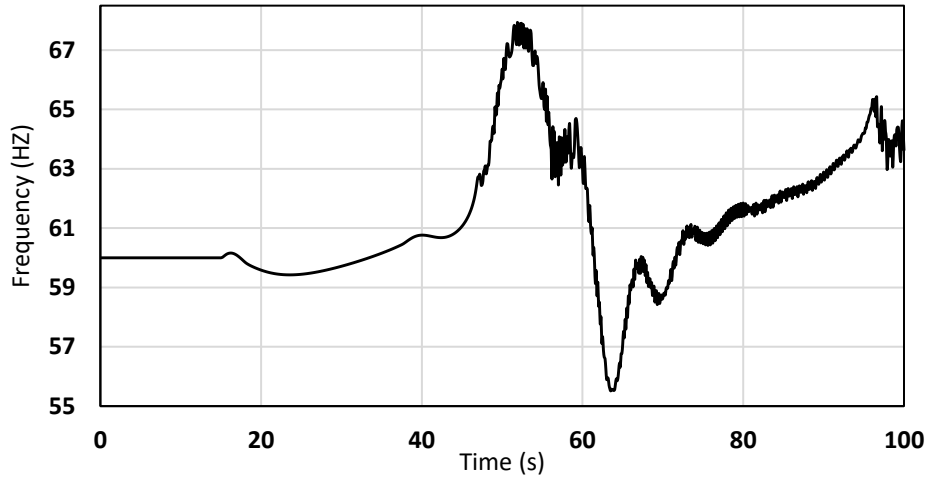

*Figure 9: 7-Bus System: Transient Frequency Response in Area 1 After a 6000 EV Attack*

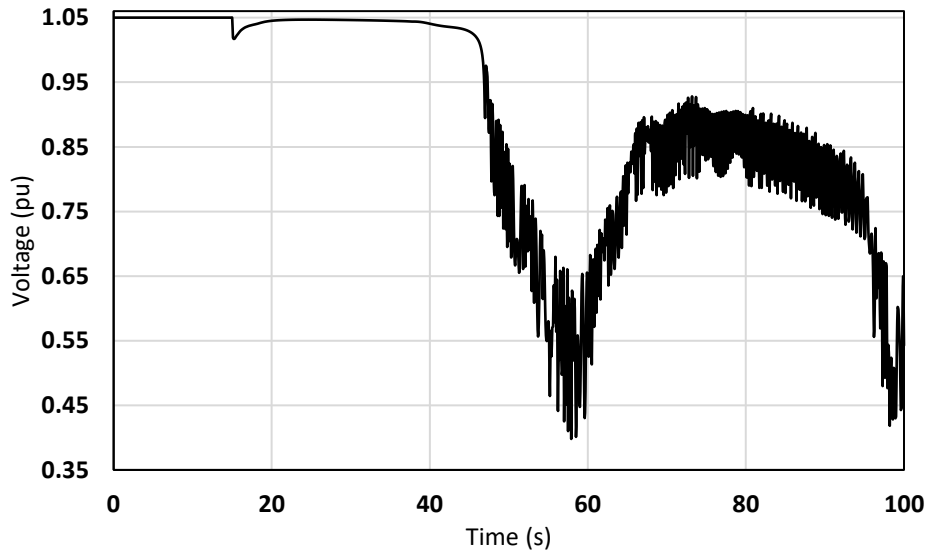

*Figure 10: 7-Bus System: Transient Voltage Response at Bus 1 After a 6000 EV Attack*

The impact of this attack is represented by the graphs in Figures 12 to 15. After the load was connected at t=15s. As depicted in Figure 12 the frequency of the top area drops after the initial attack and never recovers after that. After the initial drop in frequency caused by the extra load, the generation starts slightly being ramped up to initiate the system recovery but at around t=50s the 3 areas start oscillating at different frequencies and the generators go out of synchronization. The erratic frequency behavior and voltage fluctuation at bus 1 depicted in Figure 12 and Figure 13 respectively violate the safe operation standards and would result in tripping of the different protection relays causing large portions of the load to be dropped and possibly a blackout. One possible result of this attack would be the damage of the transformers connecting the generators to the rest of the grid as well as the transformers and other equipment interconnected within the power grid.

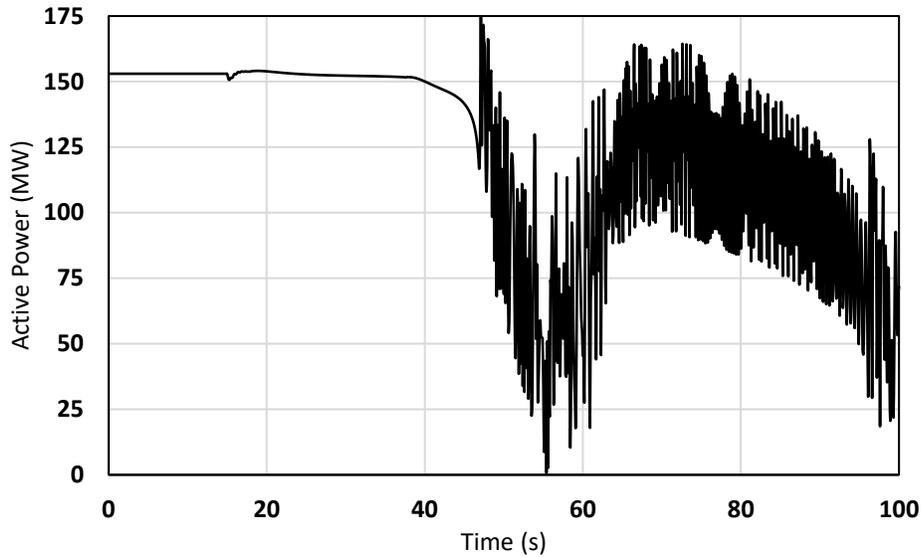

Figure 11: 7-Bus System: Generator 1 Active Power Output After a 6000 EV Attack

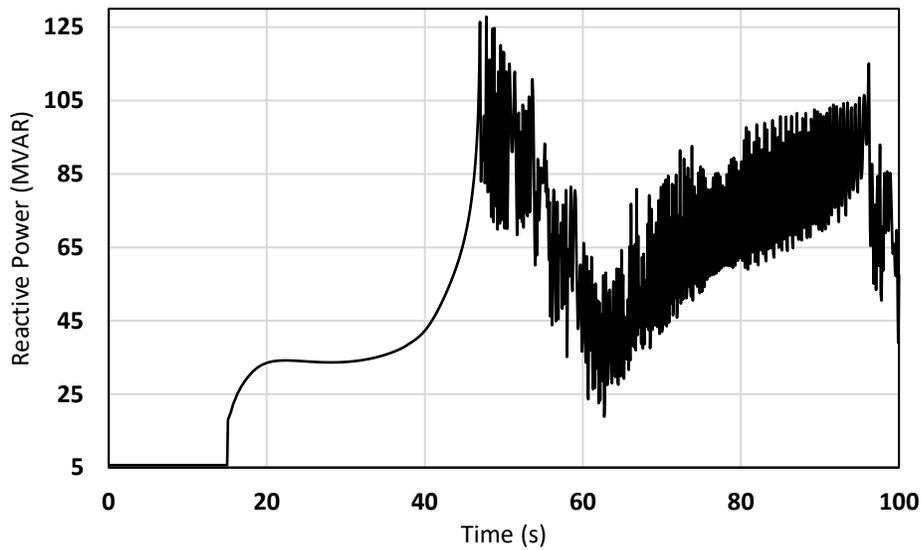

Figure 12: 7-Bus System: Generator 1 Reactive Power Output After a 6000 EV Attack

Another dangerous impact of this attack is the generator behavior in response to this sudden increase in the high VAR load. Figure 14 and Figure 15 represent the rapid fluctuation in active and reactive power output respectively. Similar to the frequency behavior, the response starts at t=15s after the attack and at around t=50s the fluctuations become erratic causing the complete destabilization of the gid and the generator protection relays to trip leading to a blackout. The most dangerous impact, however, is that the requirement to keep up with the oscillatory power requirements would result in the turbine/prime movers of the generator experiencing rapid acceleration and deceleration. This phenomenon would result in high vibration levels of the rotors leading to irreparable damage worth millions of dollars. The only way to avoid such damage is by having very sensitive protection relays that disconnect the generators before any damage occurs which would result in a blackout and the loss of millions of dollars by the affected consumers. The 3 Areas exhibit a similar frequency behavior to Figure 12, all the busses exhibit wild fluctuations similar to Figure 13. Finally, all the generators get out of synchronization and exhibit behavior similar to Figures 14 and 15.

**4.6.3   EV Load Attack – Similar MVA Load (5000 EVs)**

This second iteration of the EV attack is achieved by launching a surge in EV charging load having the same MVA load (60 MVA) as the residential load attack of Section 4.6.1. This is to demonstrate that the devastating impact achieved in Section 4.6.2 is not just due to the larger MVA load.

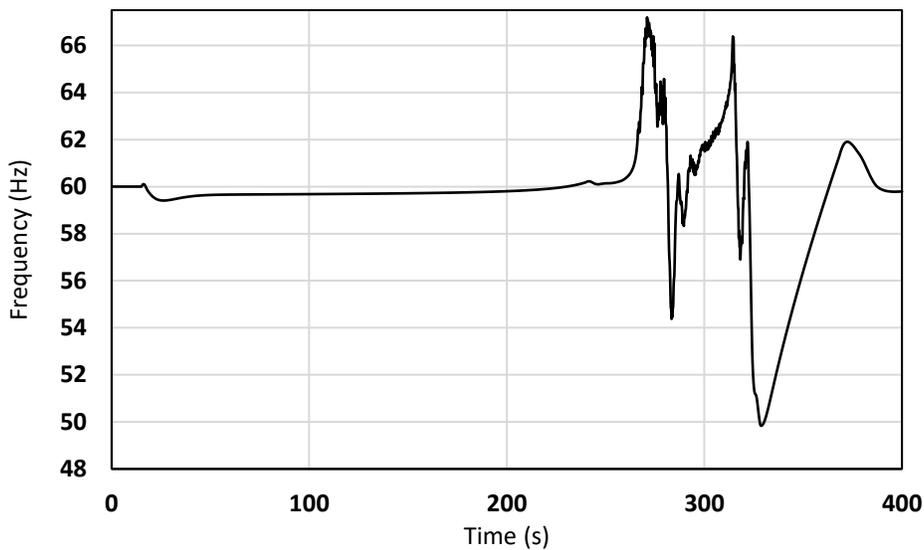

*Figure 13: 7-Bus System Transient Frequency Response of Area 1 After a 5000 EV Attack.*

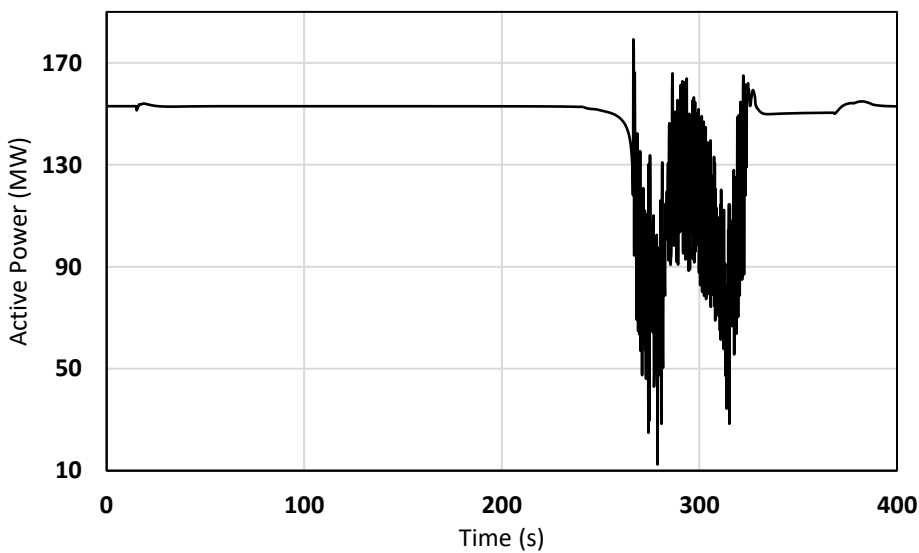

*Figure 14: 7-Bus System: Generator 1 Active Power Output After a 5000 EV Attack*

This attack is achieved by an EV charging load of 9 MW (15 MVA) at each of the 4 buses 2, 3, 4 and 5. This is equivalent to 5000 EVs or a 36 MW (60 MVA) load. As in the case of 6000 EVs, the different generators exhibit an erratic variation ion output power as well as erratic voltage and frequency oscillations. The frequency in Area 1 and generator 1 output power are presented in Figures 16 and 17 respectively. This demonstrates that even though the total apparent power of the attack is the same, the high reactive component can still destabilize the grid. The analysis of the impacts of this attack is the same in terms of causing grid instability, damage to the generators, electric equipment, consumer appliances and financial losses.

#### 4.6.4  EV Load Attack – Smaller Load (4150 EVs)

The final iteration of this attack is performed with a smaller attack in terms of EV numbers, real power load and reactive power load. The location of the compromised load used for this attack is determined by locating the most sensitive buses using the method described in Section 3.6 above. This attack is mounted through 1650 EVs or 12 MW (20 MVA) are added to bus 2, while 1250 EVs or 9 MW (15 MVA) are added to each of bus 3 and bus 5. The grid frequency, voltage, and generator output exhibit similar behavior to the 2 previous EV attacks. As with the 2 previous cases, the different generators and the 3 areas go out of synchronization, the erratic behavior results in disconnection of the generator leading to an unstable system and the system goes into a state of blackout. This serves to demonstrate that the high reactive power demand of the non-linear EV charging amplifies the impact of the attack against the power grid. A 30 MW (50MVA) EV load destabilized and damaged the grid as opposed to a 48 MW (60MVA) residential load that only caused frequency drop with no other violations and a return to stability after the generation is ramped up.

## 5  Mitigation and Detection Recommendations

The following section suggests patches for some of the vulnerabilities found in the EV ecosystem based on the industry best practices and mitigations for such weaknesses. The second part of this section describes an attack detection mechanism tailored specifically for EVs in case the vulnerability patches fail to address all weaknesses in this system.

### 5.1  Hardening the EV Ecosystem Security

The previous sections highlighted the vulnerabilities in the EV ecosystem and the attacks that can be launched against the power grid through this ecosystem. Given that these vulnerabilities are present in other IoT systems, we present practical mitigation schemes for the vulnerabilities based on well-defined solutions [46] and industry best practices. These mitigations are achieved by addressing the individual vulnerabilities in the EV ecosystem.

- Protocol vulnerabilities: to prevent the multiple vulnerabilities raised by the different protocols, a standardized set of protocols must be enforced on the EV ecosystem instead of allowing different vendors and manufactures to use their own set of protocols.

Furthermore, the optional (and sometimes nonexistent) authentication and encryption schemes must be enforced, and no data should be allowed to be transmitted in plain text.

- SQL Injection: to mitigate SQL injection attackers must be denied the chance of abusing string concatenation issues to overcome the user authentication/admin authentication process. Parametrized queries can be used to distinguish code from data thus preventing the attacker from gaining access to these accounts and denying the attacker from gaining control of the EVCS.
- XML/External Entity Injection: to mitigate XML vulnerabilities, external entities must be disabled whenever possible.
- Server-Side Request Forgery (SSRF): the IP addresses used should be validated and only pre-approved (pre-mapped) clients should be allowed to access the system.
- Cross-Site Scripting (XSS): to mitigate XSS HTTP parameters must be filtered, and user inputs must be encoded to prevent them from being manipulated by attackers.
- Comma-Separated Values (CSV) injection: to mitigate CSV injection, the system should parse the received data and reject the data that contains special characters used to trigger or execute codes
- Cross-Site Request Forgery (CSRF): to mitigate CSRF add random values to the communication process with each HTTP request to ensure the attacker cannot craft fake messages and cause system modification.
- Hard-Coded Credentials: to mitigate the issues raised by this vulnerability, the hardcoded credentials should be replaced by hash values of the credentials instead of the actual message in plain text.
- Missing Authentication: to mitigate this vulnerability, authentication should be enforced on all functionalities especially the critical functions that can be used to manipulate the charging session parameters.

Other best practices include the development of secure-by-design products instead of redesigning systems after vulnerabilities are found. Vendors and manufactures must also assess the security of their products continuously and patch vulnerabilities. Users must also be made aware of the dangers of using weak credentials to prevent attacks like the Mirai Botnet [11].

### 5.2 Attack Detection

In the following section, we propose suggestions about two detection schemes tailored to recognize EV attacks. The first suggested method tries to recognize the EV attack once it is initiated by monitoring the grid conditions by the utility. The second is a machine learning algorithm designed to monitor the usage of EV chargers by the EVCS operator or possibly the management system.

#### 5.2.1 Method 1: Attack Detection Based on Grid

The first detection method is based on an algorithm that monitors the system measurements and looks for possible EV attack properties. These properties include a sudden increase in high reactive power demand and an increase in harmonic levels. This monitoring must be done at both the transmission and distribution levels. Due to the presence of multiple VAR compensation mechanisms in the distribution network, the reactive power demand of the attack might be less visible on the transmission level. However, the states and outputs of these compensators can be monitored and incorporated into the detection algorithm as an extra indicator of the reactive power demand. Using these properties, an agent can be created to differentiate between normal system operation, system failures due to benign factors (no attack failures) and failures due to attacks initiated through the EV infrastructure.

#### 5.2.2 Method 2: Attack Detection Based on EVCS

The second detection method is based on training a machine learning model to detect the attacks on the power grid by monitoring charging stations by the EVCS operator. This agent would monitor the charging stations on two levels that work as part of the same detection agent and function in a complementary fashion.

1) Level 1 Top View: Entire System Monitoring

This level of the ML agent monitors the entire collective of EVCS. This monitoring would search for inconsistent mass charging/discharging behavior. EVCS operators have historical data on the utilization of their charging stations. Inconsistencies can include large (unnatural) mass charging/discharging that deviate from the historical and anticipated behavior. Another inconsistency is the mass (unnatural) over or under-utilization of charging stations that deviate from historical data and projected usage levels. This level of the algorithm would raise alarms when abnormal behavior is detected.

2) Level 2 Station View: Individual Station Monitoring

The second level would be an agent embedded within each station (part of the firmware for example). This agent would monitor the behavior of each EV charging station. This agent would look for inconsistent charging/discharging behavior at each station. When anomalies are detected in the EVCS utilization an alarm is raised to the control center. The control center would aggregate all the alarms from all the EVCSs and incorporate them into its decision-making process. The attacker might be able to manipulate the entire botnet of compromised EVs to craft an attack that remains stealthy to the operator. However, each charging station can detect deviations from its usual utilization habits. Even if the alarm raised by a charging station is a false alarm, the operator would take decisions based on all messages received from the charging stations and not just from a single EVCS. By collecting alarms raised by all individual charging stations, the control center would be able to circumvent the stealth of the attack.

### 6  Conclusion

In this work, we gave an overview of the current EV penetration levels as well as their underlying infrastructure. We also demonstrated the presence of vulnerabilities in the different elements of the EV ecosystem that can be used by attackers to gain control over the EVCS. This control can be used to launch attacks against the power grid by inducing mass charging and discharging.

We then discussed attack scenarios through the EV ecosystem to disrupt the grid operation and demonstrated the larger impacts they can have on the grid. Given the non-linear nature of the EV charging load discussed in this paper, we compared attack simulations utilizing residential loads and EV loads and found that the EV loads can be more damaging to the operation of the grid.

The nature of the EV ecosystem also makes it an ideal target for cyber-attacks to create a botnet of high wattage high VAR load to attack the grid. We also presented how an attacker that can estimate the grid topology can utilize stability metrics to craft smarter attacks using a smaller number of EVs while maximizing the impact. Finally, we presented patches for the vulnerabilities using industry best practices and suggested two possible detection strategies against EV attacks.